\documentclass[preprint,12pt]{elsarticle}

\usepackage{graphicx}

\usepackage{amssymb}
\usepackage[acronym]{glossaries}
\makeglossaries

\newacronym{HP}{HP}{Hewlett Packard}
\newacronym{tio2}{TiO$_2$}{Titanium Dioxide}
\newacronym{al2o3}{Al$_2$O$_3$}{Aluminium Oxide}
\newacronym{reram}{ReRAM}{Resistive Random Access Memory}
\newacronym{ups}{UPS}{Uni-Polar Switching}
\newacronym{bps}{BPS}{Bi-Polar Switching}
\newacronym{eels}{EELS}{Election Energy-Loss Spectroscopy}

\newcommand{\D}{$^{\circ}$}

\usepackage[nodots]{numcompress}

\journal{Materials Chemistry and Physics}

\begin{document}

\begin{frontmatter}



\title{Drop-coated Titanium Dioxide Memristors}


\author[uwe]{Ella Gale\corref{cor1}}
\ead{ella.gale@uwe.ac.uk, +44 117 32 82461}
\author[uwe]{Richard Mayne}
\author[uwe]{Andrew Adamatzky}
\author[uwe]{Ben de Lacy Costello}

\cortext[cor1]{Corresponding Author}

\address[uwe]{Unconventional Computing Group, Frenchay Campus, University of the West of England, Bristol, UK, BS16 1QY}

\begin{abstract}
The fabrication of memristors by drop-coating sol-gel Ti(OH)$_4$ solution onto either aluminium foil or sputter-coated aluminium on plastic is presented. The gel layer is thick, 37$\mu$m, but both devices exhibit good memristance I-V profiles. The drop coated aluminium foil memristors compare favourably with the sputter-coated ones, demonstrating an expansion in the accessibility of memristor fabrication. A comparison between aluminium and gold for use as the sputter-coated electrodes shows that aluminium is the better choice as using gold leads to device failure. The devices do not require a forming step.
\end{abstract}


\end{frontmatter}


\section{Introduction}

Memristors have been credited with the possibility of revolutionising many areas of computational science such as memory~\cite{15} and neuromorphic computation\cite{71,51}. Since the announcement of the first documented two-terminal memristor~\cite{15} (the first three-terminal memristor~\cite{29} having been made contemporary with Chua's theoretical prediction~\cite{14}) researchers have been eager to experiment with memristors, but they are difficult to synthesize and not yet commercially available. An important break-through in this area was the announcement of a solution processed memristor~\cite{28}. Although this memristor used the same `memristive' material as \gls{HP}'s nanoscale memristor, \gls{tio2}, the electrode material was aluminium rather than platinum. The authors stated that the aluminium did not have an effect on the mechanism because switching was also seen with gold electrodes. However, another recently announced memristor device~\cite{101} with aluminium electrodes (with a graphene oxide substrate rather than \gls{tio2}) has been shown to have different I-V characteristics if gold is used as an electrode. Similarly including \gls{al2o3} in gold electrode and \gls{tio2} junctions was found to promote hysteresis~\cite{123}. Finally, it has been stated~\cite{15} and widely accepted that memristors should be nanoscale devices due to reasoning based on the device thickness term in equations presented in~\cite{15}.

\gls{reram} is a field closely associated with memristors. Its actual relation is controversial as it has been claimed that all resistive switching memories are memristors~\cite{119}, this would implicitly include \gls{reram} devices, and similarly, it has been claimed that true memristors do not exist and reported memristors are actually \gls{reram}. We offer no opinion on this, but instead aim to discuss evidence from the \gls{reram} field that \gls{al2o3} may be involved in \gls{tio2} memristors. 

In the field of \gls{reram} there are two types of switching: \gls{ups} and \gls{bps}~\cite{155}. \gls{bps} closely resembles Chua's memristor plots, whereas \gls{ups} involves a much more definite jump in resistance values, usually an order of magnitude at least, although it still fits the definition for memristance. The memristive switching reported in~\cite{28} resembled \gls{ups} in that it has a large jump in resistance values. Both \gls{bps} and \gls{ups} have been reported in Pt/\gls{tio2}/Pt electrodes~\cite{157} and resistive switching has been recorded in \gls{tio2} thin films grown by atomic layer deposition~\cite{159}. It might seem strange to attempt to compare different memristors/\gls{reram} devices made of similar materials fabricated in different ways, but Magn\'{e}li phases, a reduced-oxygen-content type of \gls{tio2}, have been recorded in conduction filaments (widely believed to be the cause of switching in \gls{reram}, see for example~\cite{165}) in \gls{reram} devices and is implicated in memristor operation~\cite{130,154}.

It is known that \gls{al2o3} thin films undergo \gls{ups}~\cite{179} and that Al/anodized Al/Al devices can undergo resistive switching without needing a forming step. Al/\gls{al2o3} based devices can even be fabricated on a flexible plastic substrate~\cite{171}. \gls{al2o3} is implicated as being involved in \gls{tio2}-based \gls{reram} switching as \gls{eels} of a Al/\gls{tio2}/Al based resistive memory confirmed the presence of \gls{al2o3}~\cite{174} and furthermore adding extra \gls{al2o3} improved the operation of Al/\gls{tio2}/Al memory~\cite{189}.

In this paper, we present the creation of drop-coated Al/\gls{tio2}/Al memristors, demonstrate that they undergo memristive \gls{bps} and compare results with aluminium and gold electrodes to elucidate whether \gls{al2o3} might be involved in their operation. These devices can be synthesised with equipment available in a standard chemistry lab, simplifying the methodology still further and widening the field of researchers who can experiment with memristors.

\section{Material and Methods}

Sol-gel preparation based on~\cite{28,96}. A three-necked flask was set-up to distill under flow of dried nitrogen, then glassware was pre-heated to 120\D C to remove water. 5ml of titanium(IV)Isopropoxide 99.999\%, 20ml 2-methoxyethanol 99.9\% and 2ml ethanolamine 99+\% were injected into the flask in that order, the mixture was then stirred for an hour at three temperatures, room temperature, 80\D C, 120\D C, before the resulting blood red solution in the reaction vessel was allowed to cool to room temperature. 10ml of dry methanol was injected and the nitrogen flow turned off, after the vessel was filled with a positive nitrogen atmosphere, stoppered and left overnight to form a colourless Ti(OH)$_{4}$ (sol). A further 10ml of methanol was injected to prevent atmospheric water from reacting with the sol. This was then further diluted 1:50 in dry methanol. For the aluminium substrate comparison aluminium electrodes were sputter-coated onto PET plastic. 

\begin{figure}[!t]
\centering
\includegraphics[scale=2]{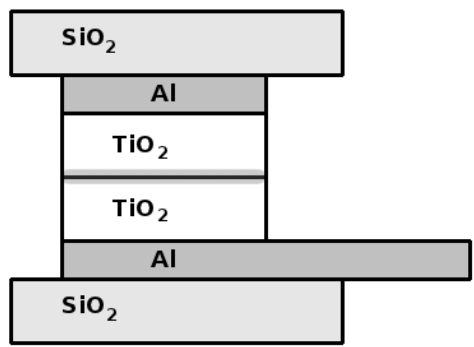}%
\caption{Device schematic. Crossed aluminium electrodes mounted on glass drop-coated with $a$-TiO$_2$ sol-gel and sandwiched together.}
\label{fig:scheme}
\end{figure}

The drop-coated memristors were fabricated using two different methods. For the simplest, two glass substrates were first covered in aluminium tape, with excess tape folded over and overhanging the edge of the glass to allow connections to the memristor. The 1:50 drop-coating solution was applied and left for half an hour in a clean fume hood (ambient air), until the white TiO$_{2-x}$ gel layer was visible, before a second drop was added and allowed to dry. The uncoated aluminium tape was then cut away and removed from the glass, except for a narrow strip which acts as a connection to the aluminium-tape overhang. Both sides were given yet another drop of Ti(OH)$_{4}$(sol) and as soon as the methanol had evapourated the two substrates were assembled as a sandwich and taped together. The best results were achieved when the two substrates had the aluminium tape `wire' at 90 degrees to each other so the only place the two electrodes were closer to each other was where the sol had been deposited. The entire drop-coating process was completed within 1 hour, the time Ti(OH)$_{4}$(sol) takes to convert to TiO$_{2-x}$(gel)~\cite{95}. To get a better aluminium surface, previously sputter-coated plastic was cut to shape, stuck to a glass substrate and then coated as above. In all devices the back of the glass substrate was covered in masking tape to prevent the measurement of glass surface effects at very low currents (10$^{-11}$A). Devices were left overnight to dry prior to measurement. Annealing the TiO$_{2-x}$ (as suggested in~\cite{96}) was found to cause short-circuits. The general scheme of the device structure is shown in Figure~\ref{fig:scheme}.

An approximate layer thickness calculated by preparing 14 double drops (as for assembling half the device) on aluminium tape covered glass surface that was previously weighed, allowed to dry in ambient air and weighed again. The actual layer thickness was verified by deconstructing a used device, slicing through it with a razor blade and using a Phillips XL-30 ESEM (Environmental Scanning Electron Microscope) to image it.

To investigate the TiO$_2$ layer, X-ray Photoelectron Spectroscopy, (XPS) and X-ray diffraction analysis (XRD) were carried out. XPS analysis was performed with a Thermo Fisher Scientific (East Grinstead, UK) Escascope equipped with a dual anode X-ray source (AlK$\alpha$ 1486.6eV and MgK$\alpha$ 1253.6eV). Samples were analysed under high vacuum ($<5\times10^{-8}$mbar) with AlK$\alpha$ radiation at 250W (12.5kV; 20mA). Following the acquisition of survey spectra over a wide binding energy range, the C1s, O1s, Al2p, Si2p and Ti2p spectral regions were then scanned at a higher energy resolution such that valence state determinations could be made for each element. XPS data analysis was carried out using Pisces software (Dayta Systems, Bristol, UK) with binding energy values of the recorded lines referenced to the adventitious hydrocarbon C1s peak at 284.8eV. The XRD was performed with a Philips Xpert Pro diffractometer with a CuK$\alpha$ radiation source($\lambda$=1.5406\AA) was used for XRD analysis (generator voltage of 40keV; tube emission current of 30mA). XRD spectra were acquired between 2$\theta$ of 10-90\D, with a step size of 0.02\D and a 1s dwell time. Phase identification was then performed using the ICDD (formerly JCPDS) spectral library and standard curve fitting software.

To elucidate the effect of aluminium, two types of memristors were made, those with two gold sputtered coated plastic electrodes and those with one gold-sputtered and one aluminium sputtered plastic electrode.

The memristors were measured on a Keithley 617 programmable electrometer which allowed the measurement of currents from pA-3.5mA. Measurements were performed with a voltage step size of 0.05V, a triangular voltage waveform and a measurement rate of 1s or 0.5s: this is the D.C. equivalent to an A.C. voltage frequency of 1mHz or 2mHz. To measure the time window of the current response, the devices were written for 200s, and then read every 19s.

\section{Results and Discussion}

\subsection{Material Properties}

\begin{figure}[!t]
	\centering
		\includegraphics[scale=0.5,keepaspectratio=true]{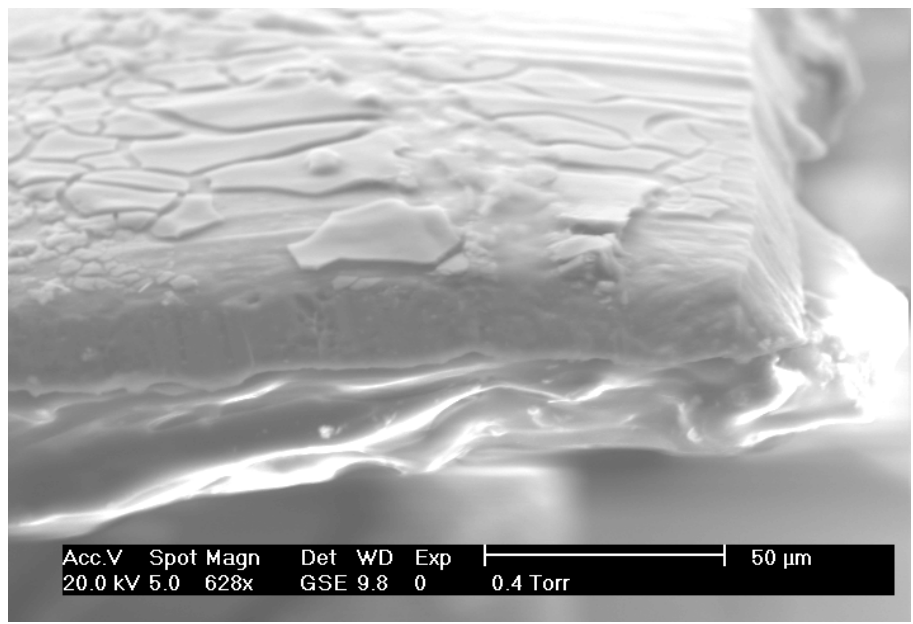}
	\caption{SEM micrograph of a slice through a half of one of the used devices showing the aluminium electrode, the thickness of the film and the flaking from the film on the half of the device. }
	\label{fig:SEM1}
\end{figure}

Figure~\ref{fig:SEM1} shows an SEM micrograph of a device that has been deconstructed. The gel layer is relatively amorphous in form (there are no visible layers from the 2 drops) and the `islands' on the top likely part of the interface between the two gel layers. The lighter substrate below the gel is the aluminium foil. 

Drop coating for 14 half devices deposited 4.9mg of TiO$_2$ sol-gel material, which assuming a stoichiometric ratio and a density similar to anatase, gives an approximate thickness of 2.54$\mu$m and thus an approximate device thickness of 5.08$\mu$m. Estimates (n=21) of the device thickness taken from across the range of figure~\ref{fig:SEM1} gives a mean thickness of 18.68$\mu$m (with a measurement error of $\pm0.53\mu$m and a standard error of 0.5399$\mu$m), a standard deviation of 2.47$\mu$m (the gel layer is thicker towards the centre of the drop which is the left hand side of figure~\ref{fig:SEM1}) and this gives an approximate gel layer thickness for an assembled device of 37.4$\mu$m. The measured device thickness is 7.4 times bigger than the thickness estimated from the deposited mass, indicating that the density of the amorphous TiO$_2$ in our devices is approximately 7.4 times less dense than anatase (and 8.2 times less dense than rutile). 

This TiO$_2$ sol-gel layer is three orders of magnitude thicker than the nanoscale thin films in~\cite{15,28,M1c} and this demonstrates that memristance can be found in micron-thick memristor systems and contradicts the assertion in~\cite{15} that memristors must have nanoscale thicknesses.

\begin{figure}[!t]
	\centering
		\includegraphics[scale=0.5,keepaspectratio=true]{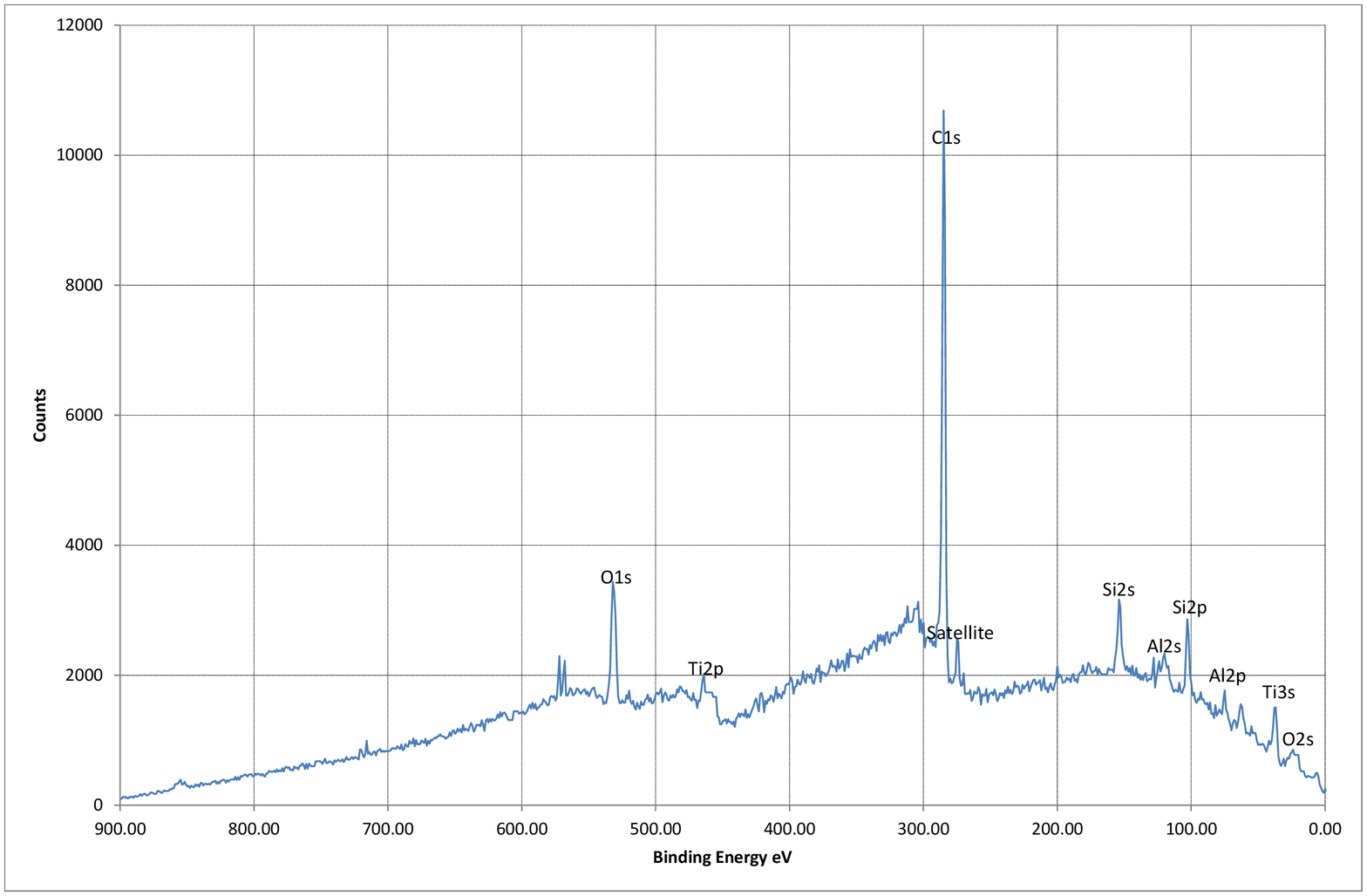}
	\caption{XPS Spectra of the inner surface layer of an opened-up device. The sample shows silica from the glass substrate, aluminium from the electrodes and advantitious hydrocarbon. There are no peaks for Ti(OH)$_4$, TiO, rutile or anatase. The peak at 458.7eV is within the range of TiO$_2$ and suggests that we have amorphous TiO$_2$.}
	\label{fig:XPS}
\end{figure}

To further investigate the qualities of the TiO$_2$ sol-gel layer both XPS and XRD analyses were carried out. Figure~\ref{fig:XPS} shows the XPS spectra of the TiO$_2$ sol-gel. In addition to the sol-gel, the XPS showed Si 2p peaks at around 100eV~\cite{269} for silica (11.6\% by atomic weight), due to the glass substrate, and Al 2p peaks (7.4\%) at 75.5eV from Al$_2$O$_3$ on the electrodes (Al metal would appear at $\approx$72.9eV). The oxygen 1s peak (7.3\%) was a mix of two peaks, one at 532.5eV for silica and one at 530.6 due to both Al$_2$O$_3$ and TiO$_2$. The Ti peak contributed 3.2\% and was at 458.7eV. (The C 1s peak is due to advantitious hydrocarbon). The Ti 2p$_{3}$ peak was found at 458.7eV which is within the general range for TiO$_2$ and correlates with the 458.9eV peak seen before~\cite{268} for amorphous TiO$_2$ sol-gels. 

\begin{figure}[!t]
	\centering
		\includegraphics[scale=0.75,keepaspectratio=true]{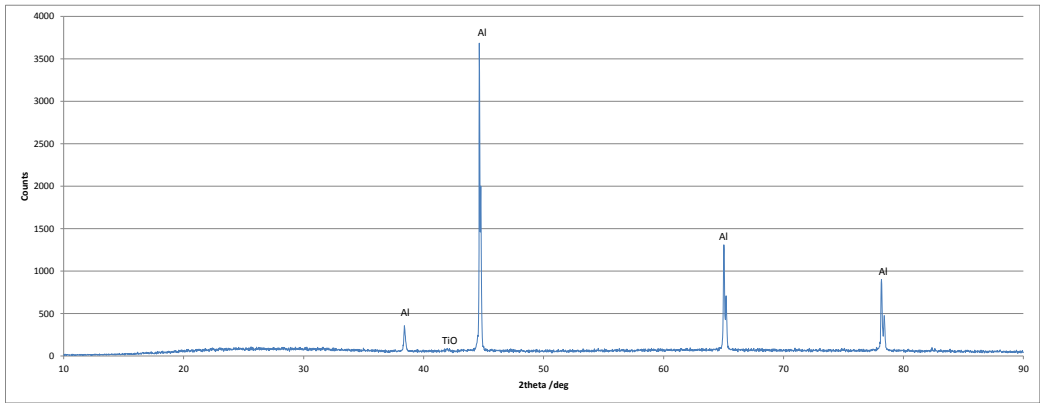}
	\caption{XRD spectra of the inner surface layer of an opened-up device. There is no crystalline TiO$_2$ phase.}
	\label{fig:XRD}
\end{figure}

XRD analyses can show the presence of crystalline phases. As figure~\ref{fig:XRD} shows, the analysis software found (with high confidence) the presence of aluminium crystalline phases, which is due to the electrodes as XRD technique can penetrate on the order of microns. A small peak at 41.5\D was found that that corresponds (with low confidence) to hongquiite, a phase with TiO stoichiometry. As a hongquiite Ti peak would be found in the XPS at 435eV (rather than at the 458eV as was recorded), we believe that the TiO `peak' is actually due to local short-range order within the amorphous phase, rather than actual areas of hongquiite (note that the calculated density is also out of the range of non-stoichiometric TiO~\cite{269}). The Ti(OH)$_4$ suspension peak that might appear at 25.5\D~\cite{267} was not observed. Both anatase and rutile phases were searched for in the data and were not found by the software (with high confidence). This lack of TiO$_2$ peaks matches the amorphous TiO$_2$ sol-gel phases seen in~\cite{270}.

Taken together, the XPS and XRD results confirm that the memristors are mainly amorphous TiO$_2$, $a$-TiO$_2$. As there is no crystalline phase, it suggests that these memristors do not operate via a phase-change mechanism. This does not rule out memristance based on the growth of filaments (perhaps of an oxygen-reduced hongquiite-like phase), memristance due to the movement of interface between the two halves of the device or bulk drift of oxygen ions.

\subsection{Electronic Properties}

\begin{figure}[!t]
 \centering
 \includegraphics{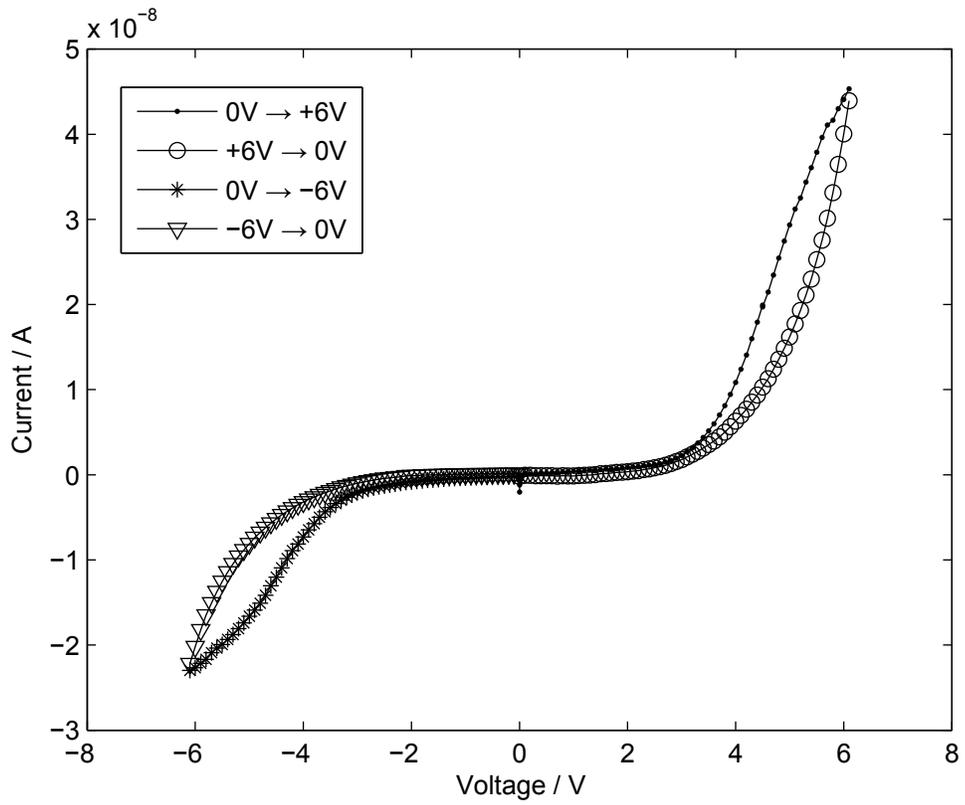}
 \caption{A macroscopic memristor made from aluminium tape. Note that the device increases resistance as it charges and has a larger positive than negative lobe. The I-V loop is pinched to zero current at zero voltage and goes around both sides of the curve in a clock-wise direction. This I-V shape is common in these memristors. This I-V curve was run with a 3 second dwell time.}
 \label{fig:S11}
\end{figure}

\begin{figure}[!t]
	\centering
		\includegraphics[scale=1,keepaspectratio=true]{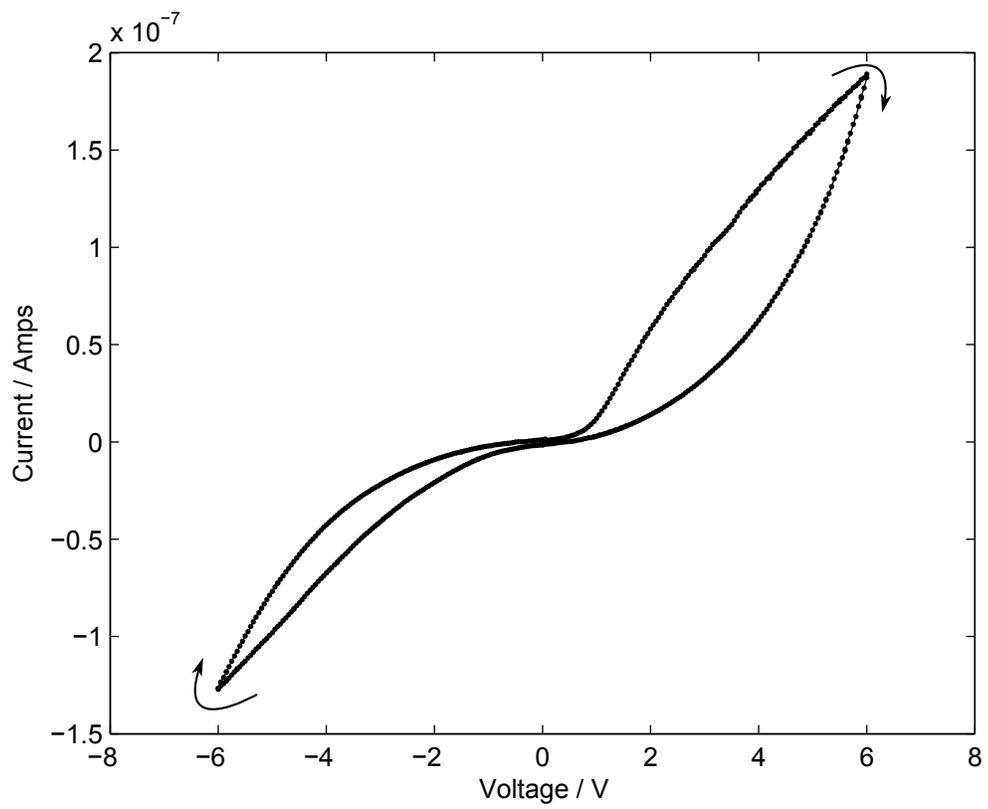}
	\caption{Typical I-V curve for a drop-coated memristor made with sputtered aluminium plastic electrodes. This device also shows the distinctive pinched hysteresis I-V curve.}
	\label{fig:Sputtered}
\end{figure}

\begin{figure}[!t]
\centering
\includegraphics[bb=0 0 478 395, scale=0.75]{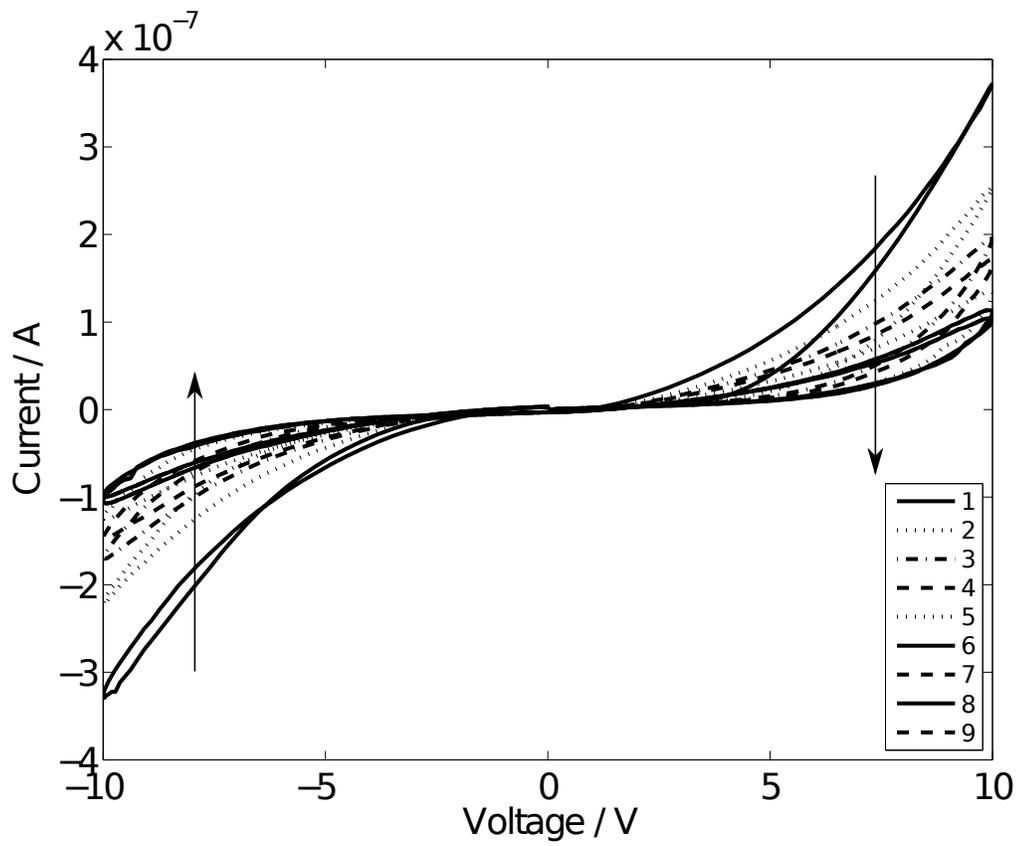}%
\caption{Reproducability on the current response to a large voltage range.}
\label{fig:Multi}
\end{figure}

Figure~\ref{fig:S11} shows a typical I-V curve for one of the drop-coated memristors prepared on aluminium tape (in fact this is from the device imaged in figures~\ref{fig:SEM1}). Those prepared on aluminium sputtered plastic were similar in form, but often had larger hysteresis, see figure~\ref{fig:Sputtered}. Both these devices start off in a highly conducting, low resistance state and switch to a low conducting, high resistance state. Both I-V loops are pinched to zero current at zero voltage and goes around both sides of the curve in a clock-wise direction. Repeated trips round the I-V loop causes the resistance to drop as the resistance change is not fully reversed over the range of the I-V loop. Figure~\ref{fig:Multi} is an example of the limits of this behaviour: the device reduces its resistance over the first 5 runs, however repeats 6 to 9 overlap more closely. Although memristors are low current devices it is  interesting to know the voltage limits that they can be taken to. Figure~\ref{fig:Multi} also shows that the drop coated memristors have a qualititatively similar behaviour if taken to a higher voltage range, such as $\pm$ 10V. 

\begin{figure}[!t]
\centering
\includegraphics[scale=0.7]{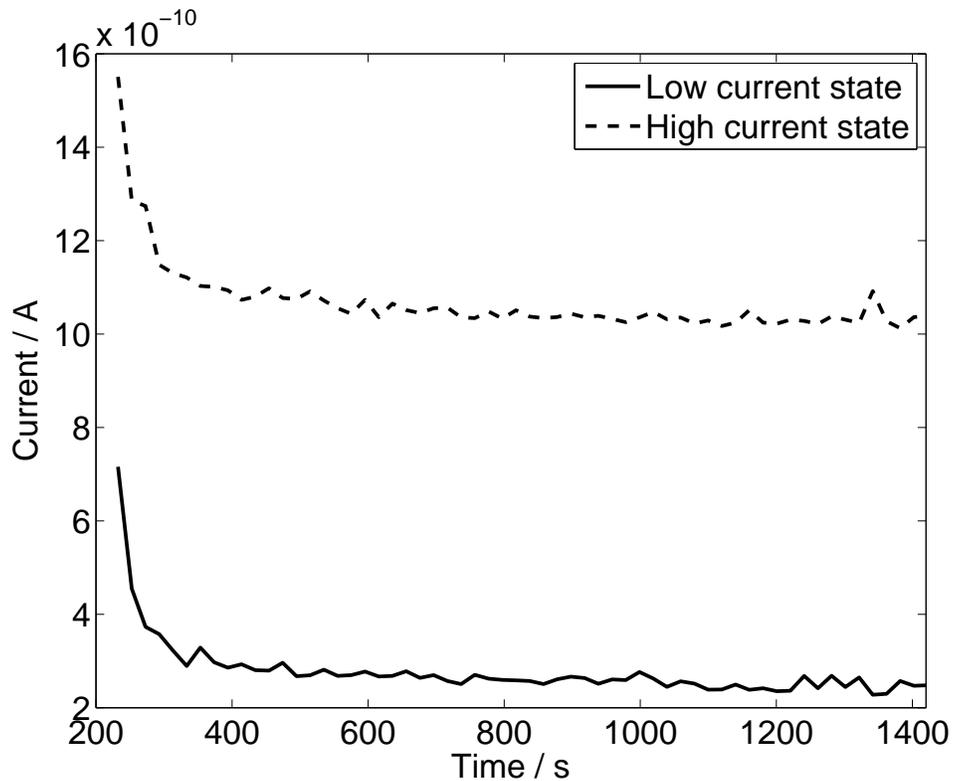}%
\caption{The time dependent response of high and low current states (after 200s of equilibration).}
\label{fig:ReRAM}
\end{figure}

Although the `fingerprint' of a memristor is its pinched I-V hysteresis loop, an often suggested use for memristors is ReRAM or general resistive memory. For this the memristor would have to hold a state. As the memristor is an analogue device, there are technically an infinite number of states the device can be set to. In Figure~\ref{fig:ReRAM} we show the time window for two such states. After the initial `writing' the device relaxes and after 200s the variation in current is minimal.

75\% (n=12) of drop coated memristors on aluminium sputtered-coated plastic were good devices (defined as possessing both a pinched I-V curve and hysteresis) with the remaining being either short-circuited or unconnected. The aluminium tape electrodes were not as flat or smooth as the sputtered aluminium electrodes, however 70\% (n=10) of the drop-coated memristors prepared on aluminium tape were good devices. This result demonstrates that good memristance I-V curves can be obtained with non-ideal aluminium surfaces. 

These devices require no forming step. We believe this is due to the fabrication method. Most of the \gls{reram} TiO$_2$ resistive memory devices were prepared by atomic layer deposition or a similarly controlled technique. TiO$_2$ memristors are believed to operate due to the movement of oxygen vacancies~\cite{15} and \gls{reram} is believed to be related to the formation of Magn\'{e}li phase filaments, therefore there needs to be an oxygen deficiency within the material. A forming step is needed to create this in vapour-deposited memristors/\gls{reram}. As drop-coated sol-gel is more amorphous, it already naturally contains areas with more oxygen vacancies, removing the need for a forming step. We expect that there are many oxygen defects in these memristors because they start in the high conducting state and then switch to a lower one under the action of applied voltage.

To elucidate the role of aluminium in the mechanism, a batch of macroscopic gold sputter-coated electrode memristors were made. Of these, one was completely short-circuited, the others were not memristors: they had tiny currents ($\sim$10 pA), straight-line profiles, no hysteresis and were comparable to the glass and aluminium electrodes test case (i.e. no semiconductor material), these are not shown in this manuscript.  

The drop-coated memristors with one aluminium and one gold electrode did not work well, and have a high failure rate: 58\% (n=12) were short-circuited. We speculate that this failure is due to either the lack of a metal oxide on the gold electrode or related to the observation that the Ti(OH)$_4$ solution was more wetting on the sputtered gold surface and thus the drops were more dispersed. Both these possibilites could lead to more short-circuited devices. Figure~\ref{fig:gold} demonstrates that the positive part of the I-V curve looks qualitatively similar to the that shown in figure~\ref{fig:S11} (as we used the drop-coating technique, the thickness of the layer is not controlled, and thus the current varies and therefore we can only compare the I-V curves between devices in a qualitative manner). There is a directionality in the devices, if the earth is connected to the gold electrode, the positive lobe of the I-V curve is larger than the negative and the negative tends to be self-crossing. Reversing the connections, rotates the curve around zero, moving the self-crossing to the negative lobe of the I-V curve (shown as an inset in figure~\ref{fig:gold}).  

As the gold electrode devices did not work and the mixed electrode devices have a higher failure rate than aluminium, this suggests that aluminium electrodes are an essential part in the operation of these drop-coated macroscopic memristors. Similar results have been seen in the nanoscale thick memristors made in our lab~\cite{M1c}. Given that the XPS showed the existence of Al$_2$O$_3$, we postulate that it is the aluminium oxide species that is important, either by facilliating the switching or by shielding the metal electrode.

\section{Conclusions}

A simpler method to make memristors for testing purposes has been demonstrated. It has been suggested that aluminium is an essential component for the operation of these sol-gel memristors. These memristors do not require a forming step. They have thicker titanium dioxide layers than those generally reported, showing that memristance does not require a nanoscale dimension within the device.

\section{Acknowledgments}
The authors would like to thank Richard Ewen for helpful advice. E.Gale is funded by the Engineering and Physical Sciences Research Council on grant EP/HO14381/1.

\section{Figures}

\clearpage

\clearpage

\clearpage

\clearpage


\begin{figure}[!t]
\centering
\includegraphics[bb=0 0 478 395, scale=0.7]{Fig3AuAlCombo.eps}%
\caption{Macroscopic memristor made with one gold sputtered electrode and one aluminium sputtered electrode, measured with the earth connected to Au. Inset shows the virgin run of another device measured with the earth connected to Al.}
\label{fig:gold}
\end{figure}



\appendix











\end{document}